\documentclass[preprint]{aastex}

\newcommand{\be}{\begin{equation}}
\newcommand{\ee}{\end{equation}}

\shorttitle{Cosmic-ray Acceleration in Sgr A*}
\shortauthors{Fatuzzo et al.}

\begin{document}

\title{Diffusive Cosmic-ray Acceleration in Sagittarius A*}

\author{M. Fatuzzo}
\affil{Physics Department, Xavier University, Cincinnati, OH 45207}
\email{fatuzzo@xavier.edu}

\author{F. Melia}
\affil{Department of Physics, The Applied Math Program, and Steward Observatory, \\
The University of Arizona, AZ 85721}
\email{fmelia@email.arizona.edu}

\begin{abstract}
Together, the Fermi-LAT and HESS have revealed the presence of an unusual
GeV-TeV source coincident with Sgr A* at the Galactic center. Its high-energy 
emission appears to be bimodal, hinting at an energizing process more 
sophisticated than mere shock acceleration. It has been suggested that 
this may be evidence of strong, rapid variability, as required if Sgr A*'s 
emission were responsible for the fluorescent X-ray echos detected in
nearby molecular clouds. In this {\it Letter}, however, we show that
stochastic acceleration in a more realistic two-phase environment 
surrounding the central black hole can accommodate Sgr A*'s
high-energy spectrum quite well. We therefore suggest that the Fermi-HESS
data alone do not necessarily provide evidence for strong variability in Sgr A*. 

\end{abstract}


\section{Introduction}
The first 25 months of observations of the Galactic center region by the
Fermi-LAT have revealed the presence of a source, 1FGL J1745.6-2900, at
energies above 10 GeV, coincident with the TeV central object HESS J1745-290
(Chernyakova et al.~2011). The combined Fermi-HESS spectrum appears to be inflected,
with a relatively steep region at intermediate energies, connected to
flatter components at both low and high energies. More specifically,
although the spectrum in the 100 MeV--300 GeV energy range can be fit
by a power law with slope $\alpha=2.212\pm0.005$ and a flux normalization
$F(100\,{\rm MeV})=(1.39\pm0.02)\times 10^{-8}$ cm$^{-2}$ s$^{-1}$ MeV$^{-1}$,
a significantly better fit is obtained with a power-law index $\alpha=2.196
\pm0.001$ between 300 MeV and 5 GeV, and a separate power law
with $\alpha=2.681\pm0.003$ in the range 5 GeV--100 GeV. The latter is also
significantly steeper than the spectrum reported by the HESS collaboration,
where $\alpha\approx 2.1$, with a total flux $1.87\pm 0.30) \times 10^{-8}$
m$^{-2}$ s$^{-1}$ above 1 TeV (Aharonian et al. 2009).

The Fermi-HESS source appears to be the high-energy counterpart to Sgr A* 
(see Melia 2007 for a review). We note, however, that although the HESS
source is coincident within $\sim30^{\prime\prime}$ of Sgr A*, its centroid
is displaced roughly $7^{\prime\prime}$ ($\sim 0.4$ pc) to the east
of the Galactic center, an issue that we shall revist below.

In earlier work (Liu et al. 2006; Ballantyne et al. 2007), we considered
the interesting possibility that the TeV $\gamma$-rays are produced via
$\pi^0$ decays generated when relativistic protons, accelerated near the
black hole, collide with hadrons farther out. Such a scenario is intriguing 
because our basic understanding of Sgr A* and its nearby environment 
precludes any possibility of this object producing a significant flux of TeV 
photons directly (Liu \& Melia 2001). However, protons can be energized 
to TeV energies by stochastic acceleration in 
a magnetically dominated funnel within 20--30 Schwarzschild radii of the 
event horizon. Then, as these cosmic
rays diffuse out through the surrounding medium, they may scatter with
hydrogen nuclei in a shocked stellar wind region and with the circumnuclear
molecular torus surrounding Sgr A*. 

But the latest $\sim 10$ GeV Fermi-LAT observations,
and more recent theoretical work with diffusive particle acceleration
(see, e.g., Fatuzzo et al. 2010; Melia \& Fatuzzo 2011; Fatuzzo \& Melia 
2012), render this basic picture inadequate for several reasons. 
First, the combined Fermi-HESS measurements show that a simple extrapolation
of the TeV spectrum into the GeV range does not provide an acceptable
explanation for the data. Second, it now appears unlikely that the
energy of a cosmic ray remains constant as it diffuses out to larger
radii. Thus, our earlier conclusion---that the incipient proton power-law index
has to be $\approx0.75$ near the black hole to produce the
observed photon index $\approx 2.25$ farther out---is not
valid when the cosmic rays undergo continued acceleration on their
way out into the circumnuclear environment.

Attempting to explain the surprising Fermi-HESS spectrum of the source
1FGL J1745.6-2900, Chernyakova et al. (2011) modified the Ballantyne
et al. (2007) treatment in several ways. Firstl, instead of adopting
a two-phase medium surrounding Sgr A*, in which the relatively cold
inner molecular torus is surrounded by a hotter, more tenuous wind-shocked
region, they assumed a single plasma with a $1/r^2$ density profile.
Secondly, they invoked a time-dependent cosmic-ray
injection process which, by virtue of diffusion-induced and energy-dependent
time delays, can produce the kind of inflected spectrum seen by the Fermi-LAT 
and HESS. They were able to produce a reasonable fit to the data, but only under
the assumption that Sgr A* is highly variable on timescales of several
hundred to several thousand years.

However, Chernyakova et al.'s (2011) analysis appears to be incomplete
and inconclusive because several important aspects of the particle
diffusion used by them are unrealistic. First and foremost, the
assumption of a constant particle energy following ejection is
difficult to justify, given what we know about the physical conditions 
surrounding Sgr A*. One of the principal goals of this {\it Letter} is
in fact to incorporate this important effect into the
calculation of the spectrum.

Second, the supposition that protons are ejected episodically by
Sgr A* is motivated by the detection of reflected X-ray emission
from the Sgr B2 cloud complex some 100 parsecs away (and other
similar clouds at various distances from the center).
This Iron-line fluorescence is often interpreted as the
light-echo of a flare produced by the supermassive black hole
hundreds (or thousands) of years ago (see, e.g., Sunyaev et al. 1993,
Fromerth, Melia, and Leahy 2001, Revnivtsev et al. 2004, Terrier et
al. 2010). But this kind of phenomenon requires a variation in the
overall power of Sgr A* by some 6 orders of magnitude during a very
short time compared to its age. Though one cannot decidedly argue
against such an unprecedented change, it is far more likely that 
the source of light responsible for these echoes was the impact 
onto the 50 km s$^{-1}$ molecular cloud behind Sgr A* by the 
supernova shell of the explosion that produced
Sgr A East (see Fryer et al. 2006).

However, Chernyakova et al. (2011) also considered a steady-state
proton ejection and were able to account for the Fermi data, albeit
under unusual cirumstances. In order for the proton diffusion to 
produce the required stratified proton energy distribution, with
the highest energy protons moving more or less rectilinearly
towards larger radii, they found that the diffusion coefficient
must be strongly dependent on the energy (i.e., $D(E)\propto
E^{0.65}$). A second principal goal of this {\it Letter} is to
calculate $D(E)$ from first principles, using the physical conditions
prevalent around Sgr A*, to see if such an energy dependence 
is realistic. As we shall see, our results do not support such a
strongly variable diffusion coefficient.

What this means, of course, is that it may be too simplistic to do 
away with the two-phase gaseous environment surrounding Sgr A*.
The physical conditions in the circumnuclear disk are drastically
different from those in the rest of the tenuous, hot medium
filling the inner 3 parsecs. Adopting an
average particle density, and a concomitantly simplified magnetic
field structure in lieu of the actual variation of these quantities
between the molecular and ionized phases, significantly alters the
particle diffusion. For these three principal reasons, it is therefore
necessary to examine whether a steady-state cosmic-ray ejection by
Sgr A* can provide a reasonable alternative explanation for its
100 MeV--100 TeV spectrum, but without having to invoke a
diffusion coefficient that depends strongly on the particle energy.

\section{Calculational Procedure}
In order to investigate how cosmic rays are
energized as they propagate through the environment
immediately surrounding Sgr A*,  we adopt
a numerical formalism based on the pioneering work of Giacalone \& Jokipii (1994)
that has been used to study the transport of
cosmic rays in static turbulent magnetic fields 
(Casse et al. 2002;  De Marco et al. 2007; Wommer et al. 2008;  Fatuzzo et al. 2010),
and has recently been extended to study the energy diffusion of cosmic-rays
in time-dependent turbulent magnetic fields in highly resistive environments.
For this latter scenario, the added presence of a turbulent electric field 
that is everywhere perpendicular to the total magnetic field drives the stochastic acceleration
of cosmic-ray particles (O'Sullivan et al. 2009; Fatuzzo \& Melia 2012).  

Following these recent works, we model the magnetic field within the inner few parsecs 
of the galaxy as a uniform field $\vec B_0 = B_0 \hat z$ superimposed by a turbulent component
(likely produced by the stellar winds in this region) that is  
expressed
as a sum of 
 $N$ randomly directed Alfv\'en waves with wavelengths $\lambda_n = 2\pi /k_n$:
\be
 \delta \vec B = \sum_{n=1}^N \vec A_n \,  e^{i( \vec k_n \cdot \vec r-\omega_n t+\beta_n)}\,,
\ee
where $k_1 = k_{min} = 2\pi/\lambda_{max}$ and $k_N = k_{max} = 2\pi/\lambda_{min}$
are, respectively, the wavenumbers corresponding to the maximum and minimum wavelengths
associated with the turbulent field, and the phase $\beta_n$ of each term
is randomly selected.  
The dispersion relations $\omega_n = v_A k_n |\cos\theta_n|$ and 
magnetic orientations $\hat A_n$ follow from linear MHD theory, where $v_A$ is the Alfv\'en speed and
$\theta_n$ is the angle between ${\vec k_n}$ and ${\vec B_0}$.  The total electric field
$\delta \vec E$ that must accompany a time-varying magnetic field
is found by invoking the MHD condition, which then guarantees that 
$\delta\vec E$ and $\vec B = \vec B_0 + \delta \vec B$ are everywhere perpendicular.
Finally, the desired spectrum of the turbulent magnetic field is set
through the appropriate choice of index $\Gamma$ in the scaling
\begin{equation}
A_n^2 = A_1 ^2\left[{k_n \over k_1} \right]^{-\Gamma}
{\Delta k_n\over \Delta k_1} 
= A_1^2\left[{k_n \over k_1} \right]^{-\Gamma+1}
\end{equation}
where $A_1$ is set by the ratio $\eta$ of the turbulent field energy
density to the underlying static field energy density (see Fatuzzo \& Melia 2012
for a more complete discussion).

In theory, the dynamics of cosmic-ray protons diffusing through the central 
few parsecs of our galaxy can be solved by numerically integrating the
governing equations of motion (see, e.g., Fatuzzo \& Melia 2012). In practice, 
however, this approach is too computationally taxing  given the large difference 
between the radius of gyration ($\sim 10^{-6}$ pc for TeV protons in the 
$\sim$ mG fields found at the galactic center) and the size of the central 
region. This point is further compounded by the large number of particles 
that must be tracked in order to obtain meaningful statistics at the highest 
energies.

In order to develop a formalism that is computationally viable,
we perform a suite of test runs 
designed to inform a simple, random-walk model 
that adequately captures the essential features of particle diffusion and 
acceleration. Specifically, we obtain displacement values
$\Delta x$, $\Delta y$, $\Delta z$ and $\Delta \gamma$ for two successive time 
intervals $\Delta t = \lambda_{max}/c$  by numerically solving the
full set of equations of motion for a particle with initial Lorentz factor $\gamma_0$
moving through a specified turbulent field environment (as defined by 
$B_0$, $v_A$, $\Gamma$, $\lambda_{max}$ and $\eta$).   The process is then 
repeated $10^3$ times for a new realization of the same field (i.e., same
field parameters but different values of the random variables $\vec k_n$ and $\beta_n$)
in order to produce a scatter plot of successive displacement pairs (e.g., $\Delta \gamma_{n+1}$
versus $\Delta \gamma_n$).   This entire process is then repeated at several different 
particle energies, and the results are used to produce correlated distributions of displacement values
that can be randomly sampled to produce a robust  random-walk model for the spatial
and energy diffusion of cosmic-rays throughout the prescribed turbulent field.  

While this process will be detailed in a later
work, we  note that diffusion coefficients
obtained  by integrating the full equations of motion for
protons with  $10^4 < \gamma< 10^8$
and a turbulent field with
$\Gamma = 5/3$, $\lambda_{max} = 0.1$ pc,
$B_0 = 1.4$ mG, $\eta = 0.55$, and $v_A = 1.1 \times 10^{-3} \, c$,
indicate that the often adopted scaling $D_\gamma \propto \gamma^{(2-\Gamma)}$
is not valid at lower energies.    Interestingly, these diffusion coefficients 
match well with those generated by our random walk model for $\gamma
\sim 10^4 - 10^5$,  with the latter scheme then yielding diffusion coefficients 
for $\gamma < 10^5$ that scale as $D_x$=$D_y$=$9.8\times 10^{24}\,\gamma^{0.11}$ cm$^2$ s$^{-1}$,
$D_z$=$1.1\times 10^{26} \gamma^{0.13}$ cm$^2$ s$^{-1}$, and $D_\gamma$=
$1.0\times 10^{-10} \gamma^{1.5}$ s$^{-1}$.

\section{Results}

Our analysis indicates that  good fits to the Fermi-HESS data can
be obtained for reasonable model parameters. For simplicity, we treat 
the inner parsecs of the galaxy as a uniform ``wind-zone"
of radius $R_{esc}$ that encompasses a high-density ``torus" 
with an inner radius of $1.2$ pc and a thickness of  1 pc.
The density distribution in the wind-zone is due almost entirely
to the interactions of stellar winds from the surrounding young stars. 
According to numerical simulations (Rockefeller et al. 2004;
also guided by Ruffert \& Melia 1994 and Falcke \& Melia 1997),
the average density in this region is $\langle n_{\rm H}^{\rm sw}\rangle
=121$ cm$^{-3}$ (outside of the clouds, defined by the condition
$n_{\rm H}^{\rm sw}<3\times 10^3$ cm$^{-3}$).  The torus is comprised 
primarily of molecular gas and has an average density $\langle 
n_{\rm H}^{cnd}\rangle=233,222$ cm$^{-3}$.
A static magnetic field $\vec B_0$ is assumed to be oriented parallel 
to the plane of the torus, with a magnitude $B_0 = 1.4$ mG,  
not unlike the  $\sim 3$ mG  value obtained by invoking 
equipartition with the hot ($kT = 1.3$ keV)  stellar-wind gas 
(Rockefeller et al. 2004).  The turbulent 
magnetic field is set by adopting the values $\Gamma =  5/3$, 
$\lambda_{max} = 0.1$ pc, and $\eta  = 0.55$.
  
Our Monte Carlo scheme is used to track the evolution of $N_p$ protons
injected at mildly relativistic energies near the black hole.  We postulate 
that these protons are drawn from the extreme high-energy tail of a 
thermal distribution near the black hole, having sufficiently large 
energies for a similar stochastic mechanism to efficiently 
accelerate them out of their thermal state.  The remaining particles 
near the black hole are then efficiently re-thermalized. We further 
postulate that none of the ambient particles in the wind zone are 
sufficiently energetic to be accelerated efficiently from their thermal state.  

The initially mildly relativistic particles random-walk through 
the surrounding medium until they either collide with
an ambient proton in the stellar-wind region  (where $\tau_{pp} = 1.5\times 10^{13}$ s), 
enter the torus, or escape 
once they diffuse beyond a radius $R_{esc}$.  The probability of scattering
in the wind-zone for a given random-walk step $\Delta t = \lambda_{max} / c$ is given by 
\begin{equation}
p = 6.7 \times 10^{-7} \left({\lambda_{max}\over 0.1\,{\rm pc}} \right)
\left( {\langle n_{\rm H}^{\rm sw}\rangle\over 121\,{\rm cm}^3}\right)
\left({ \kappa\over 0.45}\right)\left({ \sigma_{pp}\over 40 \,{\rm mb}}\right)\,,
\end{equation}
but it should be noted that particles only diffuse a distance $\sim  0.02$ pc during that time.
In contrast, the probability of scattering in the torus for a similar step-size is $\sim 1.3\times 10^{-3}$
given the much larger density.
Since a particle would need $\sim (0.5 $ pc/$0.02$ pc $)^2 = 625$ steps to traverse
the torus (assuming similar field strengths and turbulence profiles as in the wind zone), 
particles entering the torus have a high-probability of scattering.  For simplicity, we therefore assume
that any proton that enters the torus undergoes $pp$ scattering within that region.  

To estimate the escape radius $R_{esc}$ of the wind-zone, we note that $\sim 10\%$ 
of a relativistic proton's energy goes into each of the two photons produced via 
the $\pi_0$ decay following an inelastic scattering event. The turnover in the 
HESS data therefore indicates that the maximum Lorentz factor attained by the underlying 
cosmic-ray population is $\gamma_{max} \sim 10^5$. We can therefore estimate $R_{esc}$ 
by equating the acceleration time $\tau_{acc}\equiv \gamma_{max}^2/D_\gamma(\gamma_{max})$ 
$\approx 3\times 10^{12}$ s
required for particles to reach the maximum
Lorentz factor to the escape time $\tau_{esc} \equiv R_{esc}^2/D_z(\gamma_{max})$ required for particles to diffuse out of the
wind-zone region (since particles diffuse more readily in the direction parallel to the underlying field). This procedure yields an escape radius
$R_{esc} \approx 10$ pc.  While somewhat larger than the HESS source point spread 
function, this value is reasonable given the simplicity of our model. 

\begin{figure}[hp]
\figurenum{1}
{\centerline{\epsscale{0.80} \plotone{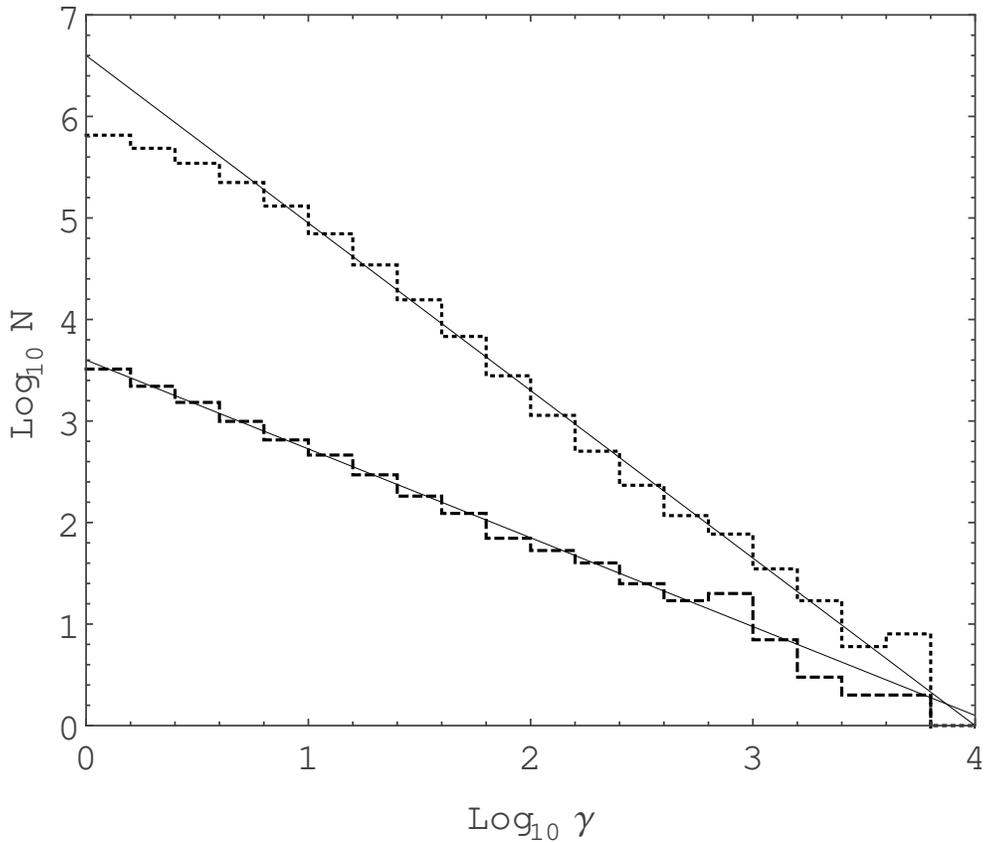} }}
\figcaption{Distributions of energy with which particles $pp$ scatter.} 
\end{figure}

A further complication for our Monte-Carlo scheme arises from the 
fact that the Fermi-HESS data range over five orders of magnitude
in energy, with the luminosity emitted at lower energies exceeding that at higher energies by $\sim$ 2 orders of
magnitude. As a result, one can estimate that there are $\sim 10^7$ times more lowest-energy
particles than highest-energy particles producing the observed emission.  Obtaining good statistics
for the entire energy range would then require that $N_p \sim 10^9$ particles be tracked.  
Since this would exceed  our computational resources, we instead track $N_p = 2 \times 10^6$ particles,
and extrapolate our results up to Lorentz factors $\gamma \sim 10^5$.

The resulting distributions in energy with which particles $pp$ scatter 
are shown in Figure 1, where the dashed histogram denotes particles that scatter
in the wind-zone and the dotted histogram denotes particles that scatter in the torus.
The solid lines show the linear fits to each distribution that are then used to extrapolate
our results up to $\gamma_{max} = 10^5$.
Figure 1 shows that the majority  of injected particles hit the torus.
This result is not surprising given that particles diffuse a greater distance along the direction of
the underlying field than across it, and are therefore naturally ``funneled" into the inner
edge of the torus for our adopted field geometry.  However, a small fraction of particles are able to
diffuse sufficiently far across the field to initially miss the torus. These particles thus diffuse throughout
the surrounding wind-zone region until they either propagate back into the torus, $pp$-scatter 
in the wind zone, or escape.  This scenario naturally produces two 
distinct power-law distributions, and a total particle distribution that
hardens from an index of $\sim 2.6$ to $\sim 1.8$ above $\sim 10^{13}$ eV.

Upon $pp$ scattering, these particles produce pions that either decay
into photons ($\pi^0\rightarrow 2\gamma$) or (in the case of $\pi^\pm$)
via the muon channels into electrons and positrons. Each particle's 
contribution to the ensuing gamma-ray emissivity (via the $\pi^0$
channel) can be calculated (see, e.g., Fatuzzo \& Melia 2003), and 
because we are assuming a steady-state ejection of particles from 
Sgr A*, the total (unnormalized) emissivity is then found by summing 
over the full ensemble of protons that scatter. The (normalized) results  
shown in Figure 2 clearly indicate that steady-state stochastic 
acceleration within the central parsecs provides a viable explanation 
for the Fermi-HESS data (the lowest energy 
data point in Figure 2 can be likely accounted for by the emission of the 
secondary leptons produced by the charged pion decays -- see for example 
Figure 1 in Fatuzzo \& Melia 2005).

A full analysis of how the model parameters affect our results is beyond the
scope of this work, and will be presented elsewhere.  We note, however, that
in general, the strength of the electric field increases as $B_0$, $v_A$ and $\eta$
increase.  As such, the particle distributions presented in Figure 1 would 
become harder as these parameters are increased.   In contrast, particle acceleration
becomes less efficient as $\lambda_{max}$ increases, and is not sensitive to 
the value of $\Gamma$ when $\eta \ga 1$ (Fatuzzo \& Melia 2012).  In addition, 
particles diffuse more readily across the underlying magnetic field as $\eta$ is increased, which has the
effect of increasing the number of protons that ``miss" the torus and 
produce the HESS spectrum.

\begin{figure}[ph]
\figurenum{2}
{\centerline{\epsscale{0.90} \plotone{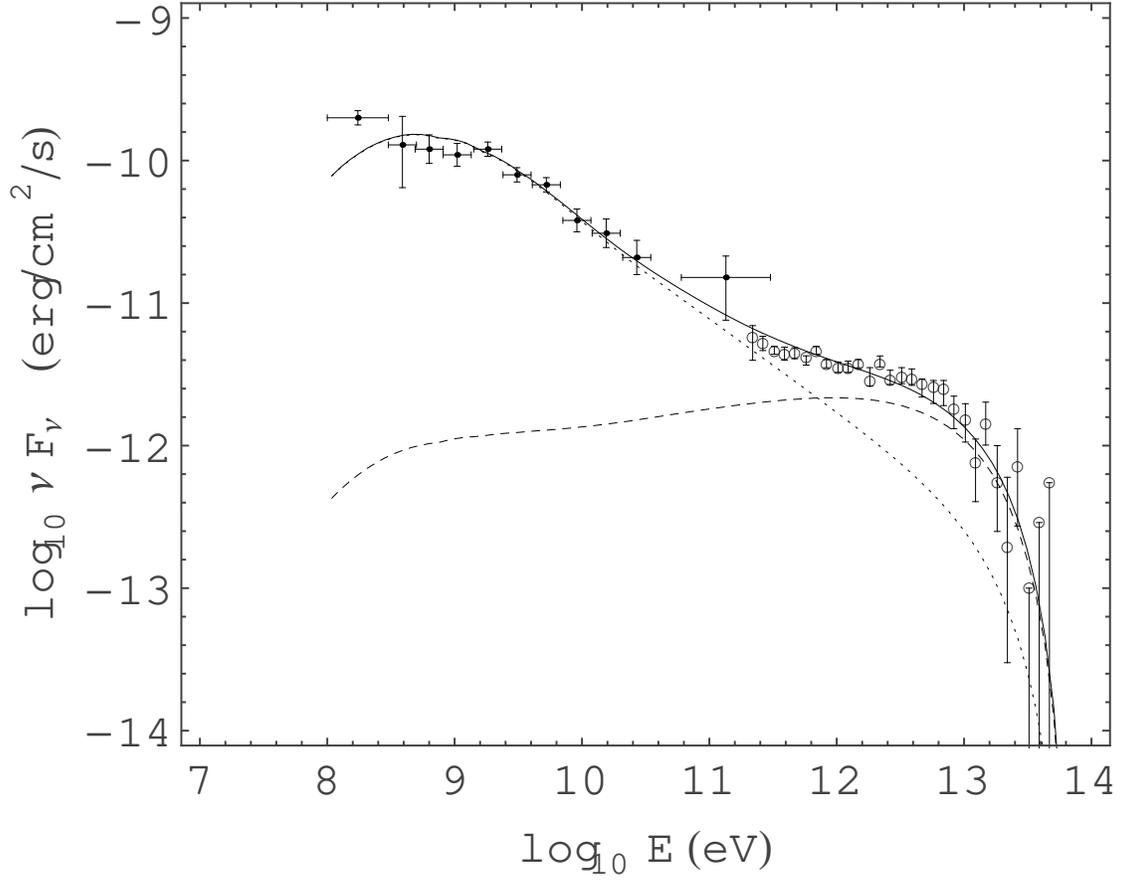} }}
\figcaption{The $\gamma$-ray emissivity (via the $\pi^0$
channel) due to particles that $pp$ scatter in the wind-zone (dashed)
and particles that $pp$ scatter in the torus (dotted) 
normalized to the shown Fermi-HESS data points.  The solid curve 
denotes the total emissivity. } 
\end{figure}

\section{Conclusions}
We have lent support to the idea that---given the right environment---cosmic
rays can be accelerated to TeV energies via stochastic processes
in a turbulent magnetic field. Although this has been suspected for many
years, we can only now demonstrate this quantitatively by comparing
detailed numerical simulations with high quality HESS and Fermi-LAT data. 
Earlier, we showed that the cosmic-ray population 
permeating the inner several hundred parsecs of the Galaxy could be accounted
for with this process, but only with the physical conditions encountered in
that region (Fatuzzo \& Melia 2012). As such, the fact that such a relativistic
particle distribution is found only near the Galactic center can be understood
as a product of the unique conditions found there. 

In this {\it Letter}, we have focused on the particle acceleration occurring 
near Sgr A* itself. The combined data exhibited in Figure~1 present quite 
a challenge to any effort at understanding how these cosmic rays are 
produced. Related to this question is the issue of whether or not these 
observations provide evidence for strong variability in Sgr A*, as 
suggested by the fluorescent X-ray echos detected in nearby molecular 
clouds.

But here we have shown that the use of a molecular torus and an interstellar 
medium filled with shocked stellar winds can provide just the right framework 
for steady-state stochastic acceleration to produce the observed two-component 
spectrum. We therefore suggest that Sgr A*'s high-energy spectrum does not 
necessarily provide evidence for the kind of rapid variability required to produce 
the fluorescent X-ray echos.

\end{document}